\documentclass[aps,prl,superscriptaddress,twocolumn]{revtex4-1}

\usepackage{amsmath,amsfonts,amssymb,amsthm,amstext,amscd,eucal}
\usepackage[all]{xy}
\usepackage{graphicx}

\usepackage{color}

\newcommand{\be}{\begin{equation}}
\newcommand{\ee}{\end{equation}}
\newcommand{\beq}{\begin{eqnarray}}
\newcommand{\eeq}{\end{eqnarray}}
\newcommand{\bea}{\begin{eqnarray}}
\newcommand{\eea}{\end{eqnarray}}

\def\[{\left [}
\def\]{\right ]}
\def\({\left (}
\def\){\right )}

\def\r2{\sqrt{2}}

\def\sst#1{{\scriptscriptstyle #1}}

\def\1{{\sst{(1)}}}

\def\CD{{\cal D}}

\def\CO{{\cal O}}

\def\CS{{\cal S}}

\newcommand{\bbibitem}[1]{\bibitem{#1}\marginpar{#1}}

\def\Label#1{\label{#1}%
  \smash{\hbox to0pt{\raise1ex\hbox{\tiny[#1]}\hss}}}
\def\noLabels{\let\Label=\label}
\def\nobbibitem{\let\bbibitem=\bibitem}

\def\Tr#1{{\rm Tr}\left(#1\right)}

\begin{document}
\noLabels 
\nobbibitem 

\preprint{UPR-T-XXXX}

\title{Helical Luttinger Liquids and Three Dimensional Black Holes}

\author{Vijay Balasubramanian}
\affiliation{David Rittenhouse Laboratory, University of Pennsylvania,  Philadelphia, PA 19104, USA}

\author{I\~{n}aki Garc\'ia-Etxebarria}
\affiliation{David Rittenhouse Laboratory, University of Pennsylvania, Philadelphia, PA 19104,  USA}

\author{Finn Larsen}
\affiliation{Michigan Center for Theoretical Physics, University of Michigan, Ann Arbor, MI 48109, USA}


\author{Joan Sim\'on}
\affiliation{School of Mathematics and Maxwell Institute for Mathematical Sciences, King's Buildings, Edinburgh EH9 3JZ, UK}


\begin{abstract}
\noindent  
Cold interacting fermions in two dimensions form exactly solvable Luttinger liquids, whose characteristic scaling exponents differ from those of conventional Fermi liquids.  We use the AdS/CFT correspondence to discuss an equivalence between a class of helical, strongly coupled Luttinger liquids and fermions propagating in the background of a 3D black hole.  The microscopic Lagrangian is explicitly known and the construction is fully embeddable in string theory. The retarded Green function 
at low temperature and energy arises from the geometry very near the black hole horizon. This   
structure is universal for all cold, charged liquids with a dual description in gravity. 
\end{abstract}

\pacs{1234}

\maketitle




At low temperatures, weakly interacting fermionic systems usually approach the well studied Fermi liquid fixed point.  The zero temperature state is characterized by a filled Fermi sea with low-energy particle-hole excitations.   Strongly interacting fermionic systems can have very different low-temperature limits with scaling exponents that are not predicted by Fermi liquid theory.   In 1+1D even a weak interaction can lead to quasi-particles with non-Fermi liquid behaviour, 
because of the restricted phase space.  
The infrared (IR) physics in this case is conformal and is generically described by a Luttinger liquid, whose  Green functions 
and scaling dimensions can be computed exactly by bosonization.   

Recent works study the physics of strongly coupled non-Fermi liquids using the AdS/CFT 
correspondence \cite{Cubrovic:2009ye,Faulkner:2009wj}: 
a fermionic operator interacts with a strongly coupled conformal field theory (CFT) that is
represented as a gravitating anti-de Sitter (AdS) spacetime with one extra dimension.  
The correlation functions of a bulk 
fermion moving in this spacetime are  related to those of the original fermion.   
A  chemical potential and temperature  are introduced in the gravitational description by 
including a charged black hole.    To understand why and how this gravitational description is 
able to model a non-Fermi liquid, an example with a solvable field theory would be helpful.  
Thus, we are led to  study the Luttinger liquid.

Here we show that a Dirac fermion propagating in the background of a 3D BTZ black 
hole \cite{Banados:1992wn} can be dual to  one component of a helical Luttinger liquid \cite{wuetal}, ie. 
a liquid where 
fermions have fixed handedness. The Fermi level is controlled by Wilson lines for a $U(1)$ 
vector potential that surrounds the black hole.  The mass of the bulk fermion controls the 
scaling dimension of the dual operator and we explicitly relate it to the effective couplings 
of the Luttinger liquid.   Our construction is embeddable in string theory, and a Lagrangian 
description  is available at weak coupling. The gravitational description is advantageous in a 
regime where the field theory is strongly coupled, 
and where properties of the liquid retain sensitivity to the UV completion.

At low temperatures the 3D black hole is nearly extremal and the analytic structure of 
the infrared Green function is controlled by the  near horizon geometry,
which can be presented as 2D AdS space with a constant electric field.  The same geometry appears near the horizon of any extremal black hole \cite{Kunduri:2007vf}, 
and controls its low-energy correlation functions. 
In this way, the nonanalytic IR behavior of every non-Fermi liquid with a gravitational dual will be related to the IR
physics of the 2D Luttinger liquid. This suggests that non-Fermi liquids in any dimension with a 
realization in gravity represent different UV completions of a universal IR 
sector.   The challenge in making this statement precise is that the different UV completions involve geometrizing different quantities in the theory, and will not be locally related to each other.

Consider the consistent truncation of Type IIB string theory to the 3D 
$SU(1,1|2) \times SU(1,1|2)$ supergravity,  
with a metric and two 
$SU(2)$ Chern-Simons  gauge fields.   The action is
$\CS = \frac{1}{16\pi\,G_N} \int d^3 x \, \sqrt{-g} \, \left( R + \frac{2}{\ell^2} \right) + \CS_{CS}(A_+) - \CS_{CS}(A_-)$ with
$
\CS_{CS} = \frac{k}{4\pi} \int \, \Tr{A\wedge dA + \frac{2}{3}\, A \wedge A \wedge A} 
$, where $k={\ell\over 4G}$ is the level of the $SU(2)$ currents. 
The vacuum solution of this theory is AdS$_3$, 
but it also has solutions consisting of the rotating BTZ black hole surrounded by Wilson lines 
\cite{Cvetic:1998xh}:
\begin{widetext}
\begin{equation}
ds^2 = -\frac{(r^2-r_+^2)(r^2-r_-^2)}{\ell^2\, r^2} \, dt^2 + \frac{r^2\, \ell^2}{(r^2-r_+^2)(r^2-r_-^2)}\, dr^2 + r^2\, \left(d\phi - \frac{r_+\,r_-}{\ell\,r^2}\, dt\right)^2 ~~~;~~~
A^3_{\mp} = \alpha_{\mp}(d\phi \pm {dt \over \ell}) \, .
\label{BTZmet}
\end{equation}
\end{widetext}
The  parameters $r_\pm$ are the outer and inner horizon radii.  Defining the left- and right- temperatures $T_{\pm} = (r_+ \pm r_-)/2\pi \ell^2$, the mass, angular momentum and temperature of the black hole are $M = (T_+^2 + T_-^2)\pi^2 \ell^2/4G$, $J= (T_+^2 - T_-^2) \pi^2 \ell^3/4G$, and $2/T = 1/T_+ + 1/T_-$. The electric 
term in the gauge fields $A^3_\pm$ is required because regularity in the (Euclidean) bulk 
imposes holomorphicity \cite{Hansen:2006wu}. The winding of the gauge fields endows the black hole with integral topological charges $Q_\pm = k\alpha_\pm$.   In the 2D 
CFT dual to AdS$_3$, this black hole is described as an ensemble of microstates with left and right Virasoro levels ${M\ell\pm J\over 2} + {k\over 4}\alpha_\pm^2$, or, in the canonical ensemble, left and right temperatures $T_\pm$.

Now consider a Dirac fermion charged under the two gauge fields propagating in this background with action
$\CS = \int\, d^{3} x \, \sqrt{-g} \, \left(i{\bar \Psi} \, \Gamma^a \CD_a \Psi - m \, {\bar \Psi} \Psi\right) 
\label{} $
where $\CD$ is a gauge covariant derivative.      According to the AdS/CFT dictionary, 
this fermion is dual to a spin-1/2 operator $\CO_m$ of {\it fixed helicity} in the 2D dual field theory \cite{Henningson:1998cd}. The operator $\CO_m$ is left handed for masses $m\ell > 0$, right handed for 
$m\ell < 0$.
Specifically if $\gamma^{0,1}$ are the $2\times 2$ 
$\gamma$-matrices in 2D and we take $\gamma^3 = \gamma^0 \gamma^1 =\sigma^3$, 
then states created by the operator $\CO_m$ are projected by 
$1 \pm \gamma^3$.  These Weyl representations are 2D analogues of fixed helicity in 4D. 

The BTZ black hole is just the $SL(2,{\mathbb R})$ group manifold, up to discrete 
identifications. This completely determines the waves propagating in the geometry.
It is then a
routine computation to take the ratios of outgoing and incoming waves at (conformal) 
infinity, with purely ingoing boundary conditions at the horizon to obtain the retarded Green function for $\CO_m$. For $m>0$, taking $\psi \propto e^{-i \omega t + i n \phi} \tilde{\psi}(r,\omega,n)$ 
and assuming non-integer $2h_\pm = m \ell + 1 \pm 1/2$,  this procedure gives  \cite{Iqbal:2009fd}:
\begin{equation}
G_R(\omega,n) = -{i\over 2} \prod_{s=\pm}  {\Gamma(1-2h_s)\Gamma(h_s - i{\omega_s\over 4\pi T_{-s}})\over (2\pi T_{-s})^{1- 2h_s}\Gamma(\tilde{h}_s-i{\omega_s\over 4\pi T_{-s}})}
\label{eq:greensfct}
\end{equation}
with $\tilde{h}_\pm = 1 - h_\pm$ and $\omega_s = \omega + s(n/\ell -2 \alpha_s)$, correcting a minor error in  \cite{Iqbal:2009fd}.  (Similar formulae with opposite conformal spin ($h_- - h_+$) follow for $m<0$ \cite{Iqbal:2009fd}.)
The  Wilson lines in (\ref{BTZmet}) shift the momenta $\omega \pm n/\ell$ by amounts proportional to $\alpha_\pm$, into which we have also absorbed the charges of the fermion under the two gauge fields.  The temperatures $T_{\pm}$ are independent for  left and right movers.    When $2h_\pm =  1,2,3,\cdots$ ($|m|\ell$ $1/2$-integral) the ratio of Gamma functions  in (\ref{eq:greensfct}) is multiplied by a factor involving di-Gamma functions ($\psi$) of the momenta:
\be
\sqrt{2}[\psi(a) - \psi(n + 1) + \gamma_E] + {1\over \sqrt{2}} [\psi(b) + \psi(b - 1)]
\ee
with $a = h_- - i \omega_-/4\pi T_+$, $b = h_+ - i \omega_+/4\pi T_-$, $n = 2h_- -1$ and $\gamma_E$ is the Euler-Mascheroni constant.  (This expression is further modified for the special case $2h_- = 1$.) The  singular normalization of (\ref{eq:greensfct}) for integer $2h_\pm$ is an artifact of neglecting these di-Gamma functions.   Below, for simplicity, we focus on the case of non-integer $2h_\pm$ although the integer values are in fact realized in the simplest string theoretic embeddings.

With $\alpha_\pm = 0$, a Fourier transform gives
$
G_R(x_+,x_-) = -i \Theta(x_+)\Theta(x_-) \left({\pi T_+  \over  \sinh\pi T_+ x_-}\right)^{2h_+}\left( {\pi T_-  \over \sinh\pi T_- x_+}\right)^{2h_-}
$
with support in the forward lightcone ($ \Theta(x_+)\Theta(x_-) = \Theta(t) \Theta(t^2 - \phi^2) $)as expected.   The overall numerical factor was determined such that  the short distance singularity (and low temperature limit) in real space takes the canonical form so that $\langle {\cal O}_m(t,\phi){\cal O}_m(0,0) \rangle = x_+^{-2h_-} x_-^{-2h_+}$.  Thus (\ref{eq:greensfct}) is the thermal Green function of an operator with spin  $h_+ - h_- = {1\over 2}$ and conformal dimension $\Delta = h_+ + h_- \geq 1$.   There is a tower of thermal poles at
$\omega_s = -i 4\pi T_{-s}(h_s + n)$ for non-negative integer $n$. These poles collapse to the real line as $T_\pm \to 0$ producing non-analytic behavior of the zero-temperature Green function $G_R(\omega,n) \propto \prod_{s=\pm} \omega_s^{2h_s - 1}$ at $\omega_s=0$, indicating the edges of the spectral bands. At zero temperature the Fermi sea is filled up to $\omega = 0$.  Thus $\omega_s = 0$ with $\omega = 0$ gives the momenta at the two edges of the Fermi surface as 
$n_\pm = 2 \alpha_\pm\,\ell$.

Using the Euler reflection formula we obtain the spectral function $4 \, A(\omega,n) = - 8 \, {\rm Im} G_R(\omega,n)$ as
\begin{equation}
\cosh\left[ \sum_{s=\pm} {\omega_s \over 4T_{-s}} \right]
\prod_{s=\pm} {(2\pi T_{-s})^{2h_s - 1} \over \Gamma(2h_s) \cos \pi h_s} \left|\Gamma\left(h_s - i{\omega_s  \over 4\pi T_{-s}}\right) \right|^2
\end{equation}
This level density is plotted in Fig.~1.   The sum and difference of the Wilson lines around the BTZ black hole ($\alpha_+ \pm \alpha_-$) move the spectral bands up/down and left/right in the $\omega-n$ plane.   The low temperature limit $T_\pm \to 0$ of the spectral function can be extracted using $  \lim_{|y|\to \infty}\frac{1}{\sqrt{2\pi}} |\Gamma(x+iy)|   e^{\frac{\pi|y|}{2}} |y|^{\frac{1}{2}-x} = 1 \, .$  Taking $\omega_\pm / T_\mp \gg 1$ this gives 
\begin{equation}
A(\omega,n) \approx 
\pi^2 \cosh(\sum_{s=\pm} {\omega_s \over 4T_{-s}})
\prod_{s=\pm} \frac{e^{-|\omega_s|/4T_{-s}} |2\pi\omega_s|^{2h_s -
  1}}{\Gamma(2h_s)\cos\pi h_s}
\end{equation}
In the region inside the spectral bands, i.e. $\omega_+ \cdot \omega_-
> 0$, the expansion of $\cosh$ gives a power law spectral density:
$A(\omega,n) \propto \prod_{s=\pm} |\omega_s|^{2h_s- 1}$.  Similarly,
outside the spectral bands ($\omega_+ \cdot \omega_- < 0$) the
spectral density vanishes exponentially: $A(\omega,n) \propto
\prod_{s=\pm} |\omega_s|^{2h_s- 1} \left(\sum_s
  e^{-|\omega_s|/2T_{-s}} \right)$, which rapidly declines with
temperature.  The structure near the edges of the spectral bands is
obtained by taking $\omega_\pm \ll T_\mp$ and using that to leading
order in small $y$
$\log|\Gamma(x+iy) / \Gamma(x)|^2 = - y^2 \sum_{n=0}^\infty (1/(x+n)^2) + \cdots$.  The right hand side defines the Hurwitz zeta function $\zeta(2,x)$.  Thus, for example, close to the spectral band boundary with $\omega_- = 0$, but with $\omega_+ \gg T_-$ we have
$A(\omega,n) \propto |\omega_+|^{2h_+ - 1} \exp\left(\omega_-/2T_+ - (\omega_-/4\pi T_+)^2 \zeta(2,h_-)\right)$.

\begin{figure}
  \label{fig:fermi-surface}
  \centering
  \includegraphics[width=2in]{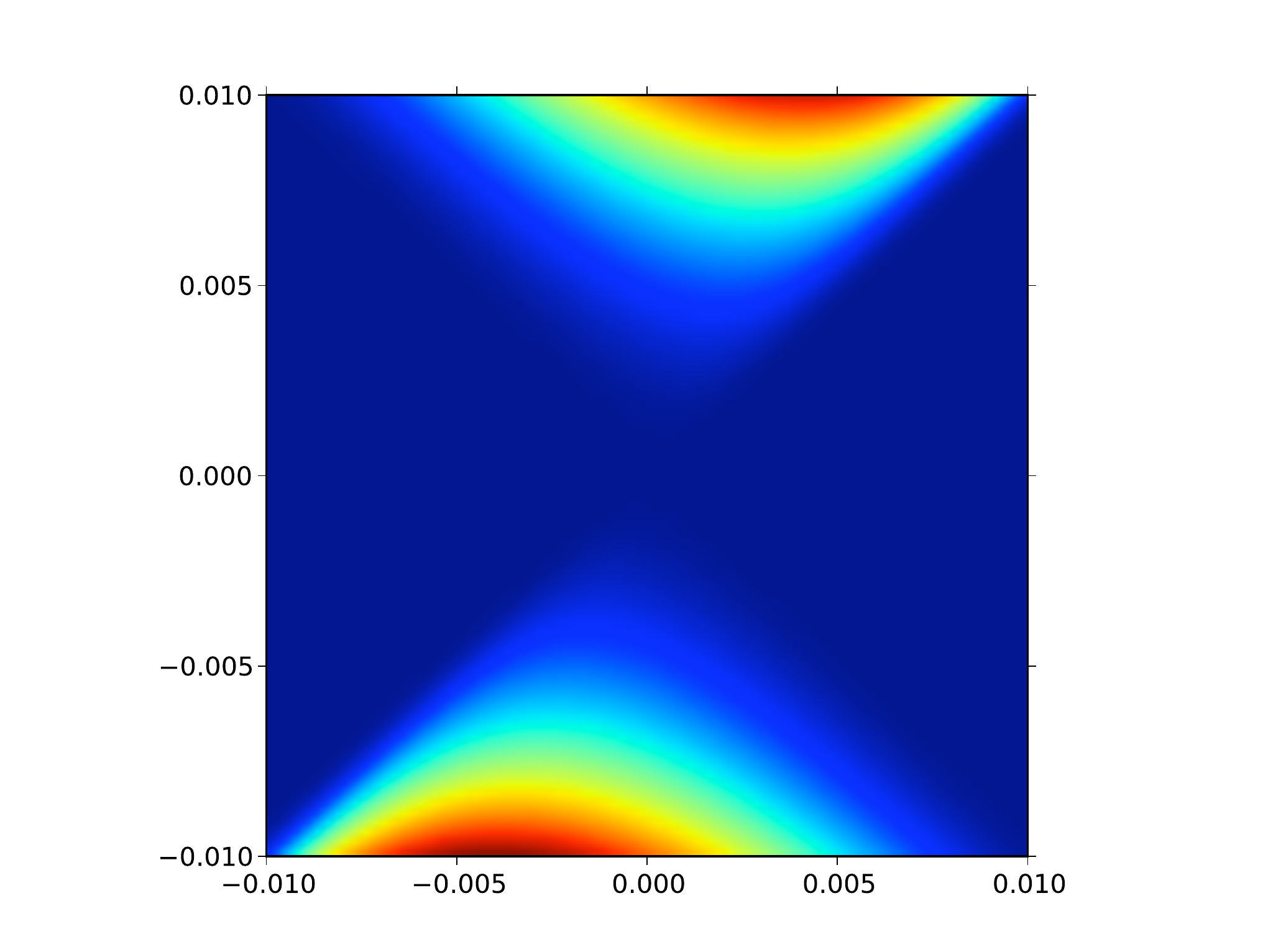}
  \caption{Spectral density for $m=1.2$, $T_\pm=10^{-4}$    and $\alpha_\pm=0$. The horizontal axis is $n$ and the vertical  $\omega$.  The spectral density vanishes rapidly outside the cones defined by $\omega \pm n = 0$.  Varying $\alpha_\pm$ shifts the spectral bands in the $\omega-n$ plane.}
\end{figure}

The system we study can readily be embedded into full-fledged
string theory, with AdS$_3$ appearing as a low energy limit, and AdS$_2$ at an
even lower energy.  In these detailed constructions (for a recent review, see \cite{Kraus:2006wn}) the fermion appears with
specific conformal weights. The simplest embedding is the $D1/D5$ system in Type IIB
string theory, with black holes that have AdS$_3 \times S^3\times T^4$ near horizon 
geometry. In this case there are fermions in chiral primary representations with
conformal weights $(h_-,h_+) = ({1\over 2}(\ell +2), {1\over 2}(\ell + 1))+{\rm c.c.}$,
$\ell=0,1,\ldots$. They have degeneracy $4$.  Thus the fermionic operator whose correlator 
we are studying is embedded in a well-defined,  UV-complete field theory -- it is a particular 
deformation of the $(4,4)$ supersymmetric $\sigma$-model on the target space $(T^4)^k/S_k$.  
While the complete field theory is strongly coupled and thus not readily solvable, it could in principle be put on a lattice and studied numerically.   

 Another standard embedding (see the review \cite{Kraus:2006wn}) is the
chiral M5-embedding in string theory, with black holes that have AdS$_3 \times S^2\times X$ 
near horizon geometry, where $X$ is a Calabi-Yau manifold. In this case the fermions in 
chiral primary representations have conformal weights of specific chirality 
$(h_-,h_+) = ({1\over 2}(\ell +2), {1\over 2}(\ell + 1))$, $\ell=0,1,\ldots$. 
Their degeneracy is $2(h_{21}+1)$, where $h_{21}$ is a Betti number of $X$. 
It is worth noting that the simplest weights are precisely half-integer, which is the case where
response functions acquire additional logarithmic behavior that is not generic. 
This is interesting but not mandatory since, going beyond chiral primaries, a discretuum of
fermion operators with spacings of order $1/k$ can also be realized in  these and 
more elaborate settings. Thus, fermionic operators with 
the properties we assume can be realized in UV-complete CFTs.

In the field theories we discuss, the fermion of interest interacts strongly with all the other 
excitations in the theory.  The collective effects of these interactions endow the fermion with an anomalous dimension.   The virtue of the AdS/CFT correspondence is that the strong interactions are conveniently resummed in this setting in terms of free propagation in a curved extra dimension.   Consider carrying out this resummation directly in the field theory at finite temperature by integrating out all the other fields.  
This will yield a complicated Lagrangian for our fermion, with
many higher order terms. However, upon running this Lagrangian down to the IR, the physics will 
be dominated by the marginal operators allowed at the interacting IR fixed point.

For spin-1/2 operators in 2D, these have been exhaustively studied (see the 
review \cite{Senechal:1999us}).  The only permitted marginal operators are
those that preserve helicity and the discrete symmetries.  To write a local interaction for a Weyl fermion we must introduce some other field.  The simplest
possibility is to assume time-reversal (TR) invariance, with a kinetic term
$
H_0 = -i \int dx \left(\psi^\dagger \partial_x \psi - \overline{\psi}^\dagger \partial_x \overline{\psi}\right)
$, 
and a four fermion dispersive interaction  coupling the two directions of motion 
\begin{equation}
H_{\rm int} = g_2 \int dx ~\psi^\dagger \psi ~\overline{\psi}^\dagger \overline{\psi}~,
\end{equation}
with spin label omitted since it is fixed by the $1 \pm \gamma^3$ projection.
This is the {\it helical Luttinger liquid}. In this realization (for which the fields exist in  the TR invariant D1/D5 theory), 
the fermions of primary interest ($\psi$) scatter off ``secondary'' fermions moving in the 
opposite direction ($\overline{\psi}$) realized in the bulk as a Dirac fermion with negative mass and 
opposite conformal spin (so the system is TR-invariant). In other realizations (including the $M5$ embedding), 
the primary fermion must interact with more general anti-holomorphic currents.

The Luttinger liquid permits an exact solution by bosonization.  The free fermion ($g_2 = 0$) is represented
as a scalar on a circle with radius $R_{\rm free}$ and then interactions are taken into account
by changing the radius to  $R^4 = {1 + {g_2/2\pi}\over 1-{g_2/2\pi}}R^4_{\rm free}$.
Interactions modify the conformal weights $(0, {1\over 2})$ of the free fermion
to $(h_-, h_- + {1\over 2})$ where 
\begin{equation}
h_- =
 {1\over 8} 
\left[ {R^2\over R^2_{\rm free}} +  {R^2_{\rm free}\over R^2}-2 \right]\simeq {g^2_2\over 32\pi^2}~.
\end{equation}
The latter approximation illustrates the small coupling behavior but the formula is exact.  Comparing with 
the formula from AdS space
\be
h_- = {|m| \ell \over 2} + {1 \over 4}  \geq {1 \over 4}~,
\ee
we get a relation between the mass of the fermion in the dual 3D gravity theory, and the coupling constant of the Luttinger liquid.   Note that the free theory ($R = R_{{\rm free}}$ or $g_2 = 0$) is never realized, since $|m| \geq 0$.

The nonanalytic structure in the low temperature Green function is due to IR physics.  
Low temperature  can be attained by taking one or both of $T_\pm \to 0$ (recall $2/T = 1/T_+ + 1/T_-$).   
A limit where only one of these temperatures goes to zero leaves the field theory in a state with 
finite chiral momentum, and corresponds in AdS$_3$ to an extremal, rotating BTZ black hole.   
The AdS/CFT correspondence reorganizes energy scales in the field theory geometrically so 
that IR physics in the field theory is associated to dynamics near the black hole horizon. Thus, 
we can extract the IR structure by examining the {\it near-horizon limit} of the bulk geometry 
and wave equations.  

The extremal ($T_-=0$)  black hole metric is
$
ds^2 = \ell^2d\eta^2 + \ell^2\,e^{2\eta}dw^+dw^- + r_+^2\,(dw^+)^2\,,
$
where $w^\pm=\phi\pm t/\ell$ and $r^2=r_+^2 + \ell^2\,e^{2\eta}$.  The near-horizon geometry can be isolated via a scaling limit $w^-\to w^-/\lambda$ and $e^{2\eta}\to\lambda\,e^{2\eta}$ as $\lambda \to 0$.  The form of the metric remains invariant in this limit, but $w^-$ effectively decompactifies, giving the ``self-dual orbifold'' of AdS$_3$ \cite{vijay-us}.  We must preserve the topological charges $Q_\pm$  associated to our Wilson lines, and that is achieved by also taking  $\alpha_+\to\lambda\alpha_+$.   
The Dirac equation is invariant in form under this scaling limit, and so the $T_-  \to 0$ Green function is
\begin{equation}
G_R = 
C \, \omega_+^{2h_+ -1} \, |\Gamma(h_-  - {i \omega_-\over 4\pi T_{+}})|^2\,\sin\pi(h_-+i{\omega_-\over 4\pi T_+})
\label{eq:greensfct-chiral}
\end{equation}
where $C$ is a temperature dependent normalization constant.  In order to match this IR 
Green function with the UV theory, we take $\lambda$ to be finite and small (rather than strictly zero).   Then 
the IR $\omega_+$ in (\ref{eq:greensfct-chiral}) is related to the UV lightcone momentum as $\lambda \, \omega_+ = \omega_+^{{\rm UV}}$, reflecting the redshift between the near-horizon and asymptotic part of the black hole geometry.
The dependence of the Green function on the chemical potential $T_+$ constitutes nontrivial 
dynamical information characterizing the system that the primary fermions interact with.

It is instructive to compare our results in $D=1+1$ to the related study of cold, 
non-Fermi liquids in $D=2+1$ \cite{Faulkner:2009wj}. The latter setting involves {\it charged} 
black holes in AdS$_4$ vs. our {\it rotating} black holes in AdS$_3$. In both studies, 
the IR dynamics (from a near-horizon limit in AdS space) is matched with the 
UV dynamics (from the asymptotic geometry), to construct the retarded Green function; 
and the crucial part of the near-horizon geometry is AdS$_2$ with an electric field.
The final result (in appendix D of \cite{Faulkner:2009wj}) for the non-analyticity
that leads to non-Fermi liquid behavior is
\begin{equation}
G^{\rm nFL}_R(\omega) = C' \, \omega^{2\nu} \, |\Gamma(\nu-iqe_2)|^2\, \sin\pi(\nu+iqe_2)
\label{4dGF}
\end{equation}
where $C'$ is a normalization constant, $q$ is the fermion charge, and $e_2$ is the electric field
(parametrizing the chemical potential).  
This precisely matches the form of (\ref{eq:greensfct-chiral}), with the identifications $T_+=(4\pi e_2)^{-1}$, $\omega_+=\omega$ and $\omega_-=q$ and  $h_-=\nu \equiv \sqrt{m^2_2R_2^2 - q^2e_2^2}$. 
Recalling that $h_- =   |m|\ell + \frac{1}{2}$, we relate the AdS$_2$ mass of
the non-Fermi liquid to the mass of our 3D fermion. 

This precise agreement between low temperature correlation functions confirms 
that the AdS$_2$ near horizon geometry is responsible in both cases for the IR behavior. The parameters
of the different UV completions are then related by comparing physical quantities in the
low energy effective theory. Despite this simple picture, the IR 
sector of the 2+1D non-Fermi liquid in \cite{Faulkner:2009wj} cannot be simply mapped into 
a Luttinger liquid. This is due to a subtle sensitivity to the UV theory. The AdS$_4$ black brane 
in \cite{Faulkner:2009wj} is charged under an auxiliary electric field, while
the electric field in our AdS$_2$ is geometrized as an extra dimension which is the direction 
of the momenta of  the Luttinger liquid.  Thus, Luttinger modes of different momenta appear in AdS$_2$ as a tower of particles with integrally spaced charges and masses, 
while the momentum modes of the 2+1D field theory in \cite{Faulkner:2009wj} appear in AdS$_2$ as a 
tower of particles with different masses, but fixed charge. These differences imply different
spectra  for $\omega_- \equiv q$ in the two cases. This feature complicates the relationship between 
the 2D physics of the Luttinger liquid and of 4D non-Fermi liquids with gravity duals.  

In summary, the IR structure of correlation functions in the holographic approach to cold Fermi liquids
always derives from the omnipresent near-horizon AdS$_2$ geometry.  
The full black hole geometry is analogous to the UV completion of an IR 
field theory \cite{Faulkner:2009wj,faulkpolch}. The BTZ black holes presented here are the most transparent UV completion.  
While convenient, our 3D completion may not capture all interesting phenomena. For example, 
a superconducting instability can usually be implemented in AdS in terms of a charged boson with a mass that 
is stable in the UV,  but tachyonic in the AdS$_2$ near horizon geometry \cite{Horowitz:2010gk}. 
An interesting feature of 
our AdS$_3$ completion is that here the stability bound is exactly the same in the UV and the IR, 
since changes in AdS radius are precisely compensated by a change in the Breitenlohner-Freedman 
bound. Thus condensation by this mechanism appears impossible. It would be interesting to understand 
in more detail what features of the UV completion drive specific  low energy phenomena in the 
holographic description of condensed matter systems.



\begin{acknowledgments}
  \noindent{\bf Acknowledgements: } We thank  S. Hartnoll, C. Kane, 
  J. McGreevy, K. Schalm, J. Teo, C. Varma, and D. Vegh for discussions, 
  and M. Rangamani for collaboration in the initial stages of the project.
  
\end{acknowledgments}

%

%
\end{document}